\documentclass[preprint,review,12pt]{elsarticle}

\usepackage{amsmath}
\usepackage{amssymb}
\usepackage{amsthm}
\usepackage{lineno}
\usepackage{nicefrac}
\usepackage{hyperref}
\usepackage{xcolor}

\journal{Extreme Mechanics Letter}

\begin{document}
\begin{frontmatter}

\title{Guided elastic waves in a highly-stretched soft plate}

\author[IL,PMMH]{Alexandre Delory}
\author[IL]{Fabrice Lemoult\corref{corresponding}}
\author[PMMH]{Antonin Eddi}
\author[IL]{Claire Prada}

\address[IL]{Institut Langevin, ESPCI Paris, Universit\'e PSL, CNRS, 75005 Paris, France}
\address[PMMH]{Physique et M\'ecanique des Milieux H\'et\'erog\`enes, CNRS, ESPCI Paris, Universit\'e PSL, Sorbonne Universit\'e, Université de Paris Cit\'e, F-75005, Paris, France}
\cortext[corresponding]{Corresponding author: fabrice.lemoult@espci.psl.eu}

\begin{abstract}
We study the propagation of guided elastic waves in a highly-stretched Ecoflex\textsuperscript{\textcopyright} plate, a nearly-incompressible elastomer. The plate is subjected to a nearly-uniaxial stress with an elongation reaching 120\% and we measure in-plane displacements of the shear horizontal mode $S\!H_0$ and of the plate mode $S_0$ coexisting in the low frequency limit. An induced anisotropy is first observed and characterized by following the phase velocities in two principal directions. Although these measurements provide an initial stress estimate, we evidence the limits of the acousto-elastic theory to predict those phase velocities in a prestressed elastomer. Taking into account the frequency dependent shear modulus of the elastomer, an experiment-driven fractional rheological model is added to the theory. This provides a proper prediction of phase velocities up to 80\% elongation.\\
\end{abstract}


\begin{highlights}
\item In-plane guided waves are measured in a highly-stretched plate of Ecoflex
\item Fundamental symmetries are broken and the induced anisotropy is fully characterized
\item Limits of the acousto-elastic theory are evidenced
\item The theory is adapted using an experiment-driven fractionnal viscoelastic model
\item The visco-hyperelastic model properly estimates complex wavenumbers up to an elongation of 80\%
\end{highlights}

\end{frontmatter}

\section*{Introduction}

The evaluation of mechanical properties is of paramount importance and the propagation of elastic waves allows to probe those properties deep inside the medium. Such an inverse problem is of fundamental relevance for geophysics imaging~\citep{shapiro_2005} or seismology~\citep{benzion_2003}. Similarly, in an industrial context, non-destructive techniques based on ultrasonic waves permit to assess the elastic constants of a material~\citep{bochud_2018,thelen_2021}.
In human body, ultrasonic waves are also used to build medical images of f\oe{}tus and other organs. Interestingly, in this specific application, the medium is considered as a fluid-like material and the images remain qualitative. Nevertheless, with the emergence of elastography~\citep{doherty_2013,sigrist_2017} in the last decades, shear elastic waves which are typical of solid-like material have revealed very useful to probe the stiffness of tissues. For example, using the acoustic radiation pressure, the Aixplorer\textsuperscript{TM} generates shear waves and track its propagation thanks to high-framerate movies, in order to map the shear velocity, thus the shear modulus~\citep{doherty_2013,gennisson_2013}. Despite some very efficient applications such as the detection of hepatic fibrosis~\citep{sandrin_2003,asbach_2010}, this technique shows some limitations for quantitative measurements~\citep{li_2017,bilston_2018,caenen_2022}.
First, due to their geometry, numerous tissues act as waveguides. The dispersive nature of guided waves in soft media~\citep{couade_2010,lanoy_PNAS_2020,delory_JASA_2022} is often not taken into account in elastography.
Second, the body is mostly made of nearly imcompressible media, which is highly deformable, and the retrieved stiffness appears to depend on applied stresses~\citep{catheline_2003,gennisson_2007,crutison_2022}. This dependence is known as the acousto-elastic effect~\citep{biot_1940,toupin_1961}. This effect is not specific to elastography but refers to the changes in elastic wave velocities with an initial stress, as fully described in~\citep{ogden_1997,destrade_2007}. In particular, many examples with analytical results are provided in~\citep{destrade_2010}. It is for instance at the basis of string instruments, where the tension is finely tuned to adjust the pitch of the musical instrument. The more deformable the medium, the more significant this effect, this is why the case of soft media is of special interest.
With the advent of highly deformable elastomers in new technologies, the class of hyperelastic media has emerged~\citep{boyce_2000,marckmann_2006}. It has been especially useful to describe the mechanical properties of biological tissues~\citep{wex_2015,chagnon_2015}.
The hyperelastic constitutive law is often considered since the velocities of shear waves directly provide a measurement for the applied stress~\citep{li_2017,destrade_2012}.

Although soft material are generally viscoelastic, the rheology is often omitted when modelling wave propagation in pre-stressed soft materials. Yet, it is well-known that rheology impacts results of usual mechanical testing. Typically, hysteresis curves will appear in tensile test experiments because the material is inherently dissipative and strain-rate dependent. For example, usual viscoelastic models are used by Cases et al.~\citep{case_2015} to describe time-dependent mechanical experiments of different soft elastomers including Ecoflex, the material used in this work. Yasar et al.~\citep{yasar_2013} probed Ecoflex viscoelastic properties using magnetic resonance elastography and found that a fractional derivative model is a better fit compared to usual viscoelastic models.
Theoretical works have been devoted to mixing the viscoelasticity and the well-established acousto-elasticity. Especially see the work on bulk waves of Destrade and Saccomandi~\citep{saccomandi_2003,saccomandi_2004,destrade_2009} or others~\citep{colonnelli_2013} where the stress tensor is rewritten as a sum of an elastic and a dissipative part. More recently, De Pascalis et al.~\citep{parnell_2019,berjamin_2022} generalises the quasi-static linear viscoelastic acoustoelasticy using memory variables. However, there is a lack of experimental validations addressing these questions.
In addition, the effect of the initial stress on the dispersion curves has always been an active field of theoretical research~\citep{ogden_1993,rogerson_1995,nolde_2004,rogerson_2009,mohabuth_2019}. Lately, Zhang et al.~\citep{zhang_2022} assess the influence of a moderate initial stress on the dispersion curves of a fractional viscoelastic plate. But again, very few experimental works address these questions. Among them, a recent study~\citep{li_2022} proposed to map stresses in thin films using the acousto-elastic effect on guided waves but without viscosity.

In this paper, we address several issues using a simple experiment. The propagation of guided elastic waves in a highly deformed (elongation reaching 120\%) soft plate is investigated experimentally and theoretically. First, the stress-induced anisotropy is observed and quantified. Such high deformations are commonly reached in tensile tests but rarely in acousto-elastic experiments. We perform systematic measurements of the guided wave velocities along with or transversely to the stress direction at different stretch ratios ($1\le\lambda_1\le2.2$). This reveals different behaviours for the two low-frequency shear horizontal and plate modes.
Thanks to the acousto-elastic theory, analytical hyperelastic predictions are built. On one side, we can recover the static stress as in~\citep{li_2022}, but on the other side, no hyperelastic model correctly describes the evolution of the modes velocities as a function of the initial elongation.
To rightfully predict those changes, we add a dissipative stress tensor to the hyperelastic stress tensor. The chosen viscoelastic model is a fractional Kelvin-Voigt model, as in~\citep{lanoy_PNAS_2020,delory_JASA_2022,yasar_2013}.
At the end, this model not only correctly predicts the changes in phase velocities of those two modes, but also the attenuation distances. Overall, by answering those multiple questions with complete experimental results, this paper connects acoustical, mechanical and medical communities.

\section*{Experimental observations}
In this first part, we want to observe elastic waves propagating in a soft plate when it is highly deformed. The experimental setup is described, as well as the method to measure the full in-plane displacement field. Two guided modes are observed and their well-known physical origins are briefly discussed. Then, the setup is used to demonstrate, and more importantly quantify, how the initial stress has induced anisotropy for those two waves and how the different velocities evolve with the stretch ratio $\lambda_1$.

\subsection*{Experimental method}
\begin{figure*}
    \centering
    \includegraphics[width=.7\textwidth]{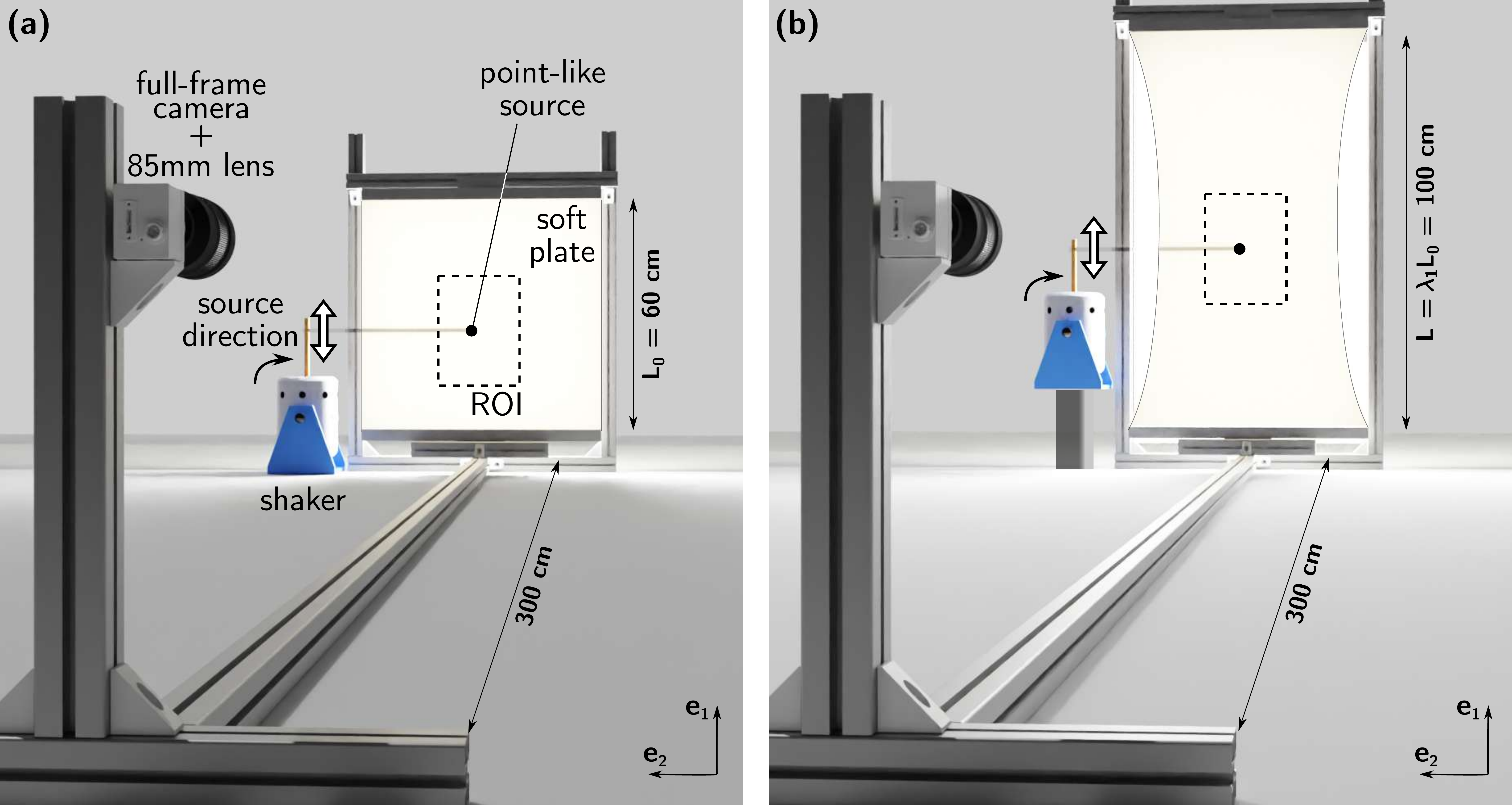}
    \caption{\textbf{Experimental setup to measure velocities in a (deformed) plate--} (a) A thin plate of Ecoflex\textsuperscript{\textregistered}~00-30 with dimensions $60$~cm~x~$60$~cm~x~$3$~mm is held in a vertical position, clamped to a frame on its top and bottom edges. Sinusoidal vibrations in the $\left(x_1,x_2\right)$ plane are generated by a shaker and extracted using a CCD camera located $3$~m away from the plate. (b) Same experimental configuration, but the frame can be adjusted to impose large deformations, reaching a stretch ratio $\lambda>2$ along the vertical axis.}
    \label{fig:setupPlate}
\end{figure*}

The experiment, presented in figure~\ref{fig:setupPlate}, has been inspired from our previous works \citep{lanoy_PNAS_2020,delory_JASA_2022}. A 3~mm thick plate made of a soft elastomer, namely Ecoflex\textsuperscript{\textregistered}~00-30, is prepared. This material is assumed to be nearly incompressible with a Young modulus $E\sim75$~kPa. The plate is held vertically and clamped at its bottom and top edges to rigid bars. This configuration allows to apply a static and large stretch in the plate.

In figure~\ref{fig:setupPlate}(a), the plate is undeformed and in a natural configuration. After applying a nearly-uniaxial stress in the plate (figure~\ref{fig:setupPlate}b), it is in a deformed configuration characterized by stretch ratios $\left(\lambda_1,\lambda_2,\lambda_3\right)$. In both configurations, a point-like source is made of two magnets fixed on both sides of the plate center. This point source is driven monochromatically with a shaker and generates elastic waves polarized in the $\left(x_1,x_2\right)$~plane. The shaker can be rotated in order to change the source oscillation direction. The excitation frequency spans a range from 50~Hz to 300~Hz. A 60-frames video is recorded using a Basler full-frame CCD camera (acA-4112-20um) mounted with a 85~mm lens. The limited acquisition frame rate of the camera is overcome using stroboscopic imaging. Each frame is then compared to a reference through a Digital Image Correlation (DIC) algorithm~\citep{DIC_Sander} and the in-plane wave field components $\left(u_1,u_2\right)$ are retrieved.

\subsection*{Guided elastic waves in a plate}
An example of acquired frame is displayed in figure~\ref{fig:first_obs}(a) for the undeformed plate. Typical displacement maps obtained when vibrating the source at 200~Hz are also shown as a colour code. Given a source vibrating along $x_1$ axis (respectively $x_2$), the displacement~$\text{Re}\left[u_1\!\left(\omega\right)\right]$ (respectively~$\text{Re}\left[u_2\!\left(\omega\right)\right]$) is displayed in figure~\ref{fig:first_obs}(b) (respectively~\ref{fig:first_obs}(c)). As a first observation one can notice that rotating the source by 90\textdegree{} involves a rotation of 90\textdegree{} of the displacement map. This demonstrates that the material is isotropic at rest and no privileged axis exists in the undeformed case. By carefully looking at the change of sign along the two main directions one can notice the existence of two distinct wavelengths, one being twice larger than the other. This effect is confirmed by applying a spatial Fourier Transform on these wave-fields. After normalization and summation in intensity of the two, the spatial spectrum of the measured waves evidences two concentric circles in figure~\ref{fig:first_obs}(d), with radii again showing this factor of 2. It corresponds to guided modes with isotropic behaviours.

Given the theoretical framework detailed in~\citep{delory_JASA_2022,royer_1999}, only three modes can propagate in this plate at this frequency: the first shear horizontal mode $S\!H_0$ and the first two Lamb modes $S_0$ and $A_0$. While $S\!H_0$ and $S_0$ are polarized in the $\left(x_1,x_2\right)$~plane, the $A_0$~mode is mainly polarized along $x_3$~axis at this frequency, and is not observed in this experiment. Furthermore, $S\!H_0$ and $S_0$ are nearly non-dispersive at this frequency, and  propagates respectively at the transverse velocity~$V_T$ and at the plate velocity~$V_P$. For an incompressible solid, it is noteworthy that $V_P = 2V_T$. All these considerations permit to label the two visible circles in figure~\ref{fig:first_obs}(d).

\subsection*{The initial stress has induced anisotropy for elastic waves in a plate}
\begin{figure*}
    \centering
    \includegraphics[width=\linewidth]{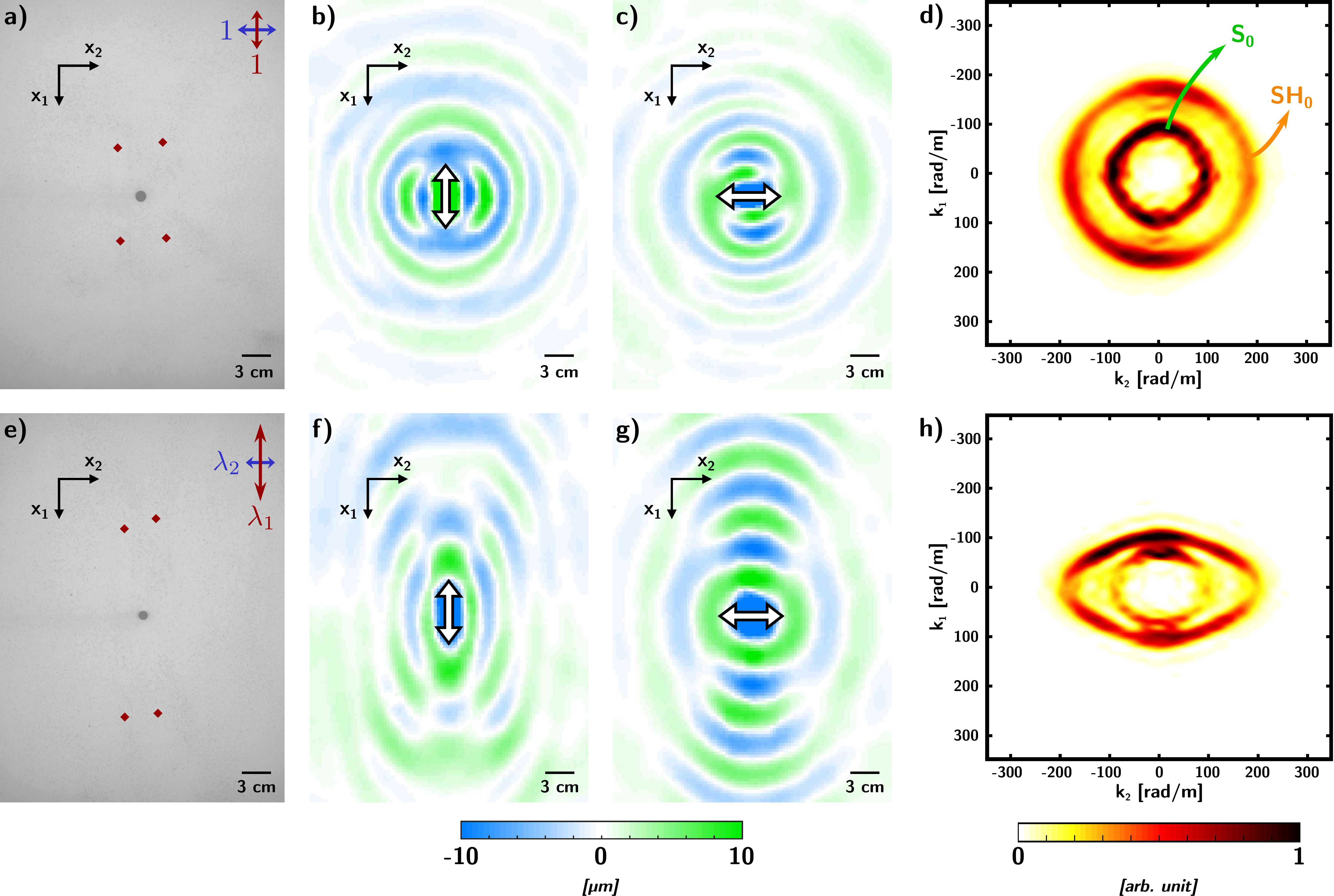}
    \caption{\textbf{Experimental 2D displacement maps in an undeformed and a deformed plate at 200~Hz --} (a)-(e) Typical pictures obtained for the initial and the deformed plate. 4 red dots are here to measure $\lambda_1 = 2.01$ and $\lambda_2 = 0.74$. (b)-(f) 2D map of the displacement $u_1$ with a source vibrating along $x_1$. (c)-(g) 2D map of the displacement along $u_2$ with a source vibrating along $x_2$. (d)-(h) Isofrequency contours for the initial and deformed plate. The spatial Fourier transforms of the two previous maps are normalized, squared and summed.}
    \label{fig:first_obs}
\end{figure*}

The same procedure is then repeated after applying a static load to the plate. The stretch ratios in directions $x_1$ and $x_2$ are measured by manually tracking the displacement of the red diamonds in figure~\ref{fig:first_obs}(e). In this specific example, the following stretch ratios are measured: $\lambda_1 = 2.01$ and $\lambda_2 = 0.74$ in the center of the plate. Here, the applied static stress is not exactly uniaxial because of boundaries but the deformation remains uniform in the central region of the plate, where the waves are monitored, as detailed in Appendix~A. We found that a great matching is $\lambda_2 = \lambda_1^{-0.41} \ne \lambda_1^{-0.5}$.

Comparing the field maps for the two different vibrating directions of figure~\ref{fig:first_obs}(f) and \ref{fig:first_obs}(g) now reveals that the system is no longer invariant by rotation: the initial deformation has induced an anisotropy to the propagation. Again, this effect is nicely caught in the spatial Fourier domain of figure~\ref{fig:first_obs}(h) where the two concentric circles are now replaced with ellipses. It appears that the elongation does not affect similarly the two types of waves since the two ellipses have different aspect ratios.

\subsection*{Quantifying the changes in phase velocities (and attenuation distances) of $S\!H_0$ and $S_0$ modes with the stretch ratio}
\begin{figure*}
    \centering
    \includegraphics[width=\linewidth]{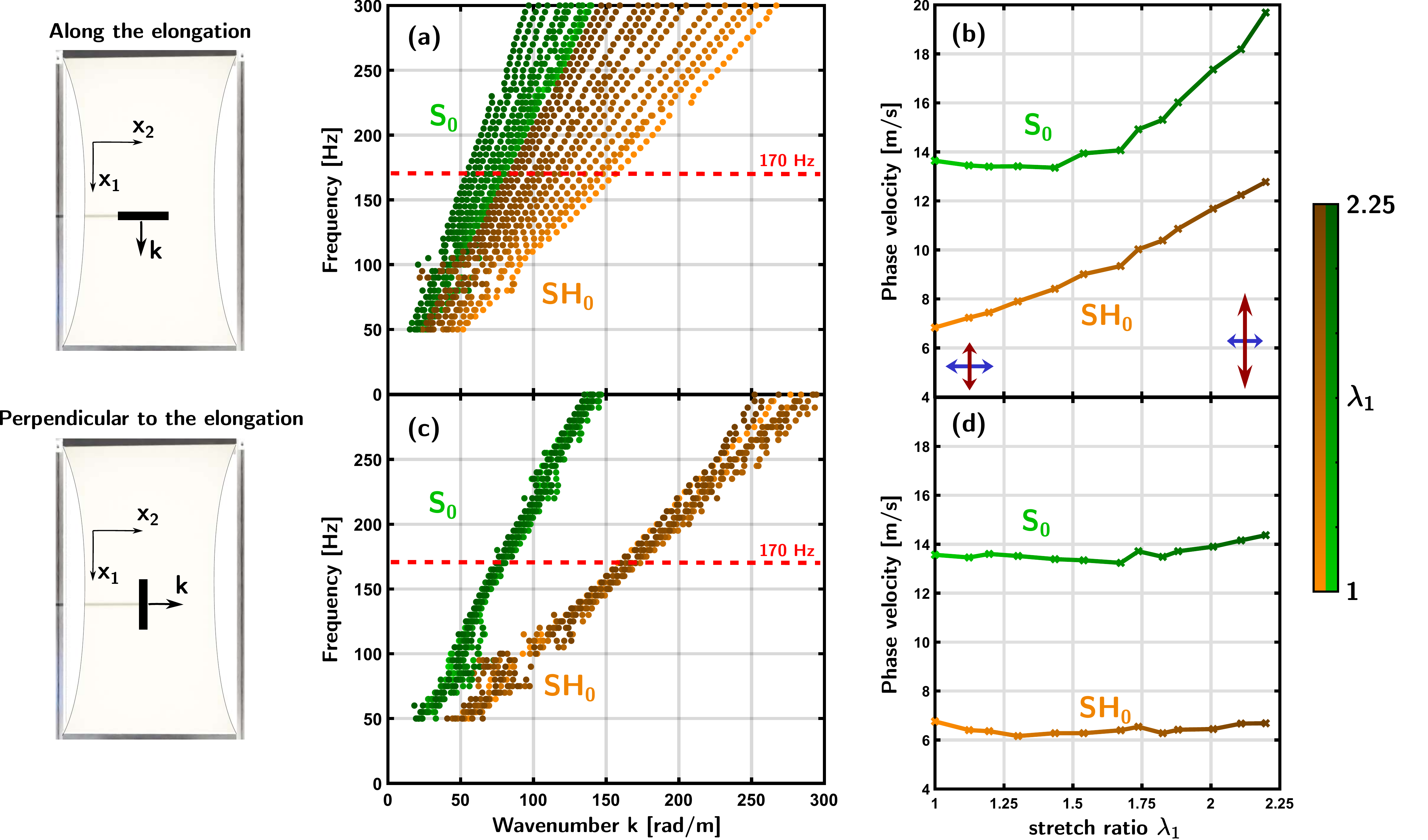}
    \caption{\textbf{Experimental dispersion curves and phase velocities at 170~Hz for a stretched plate --} The measurement in the top part (respectively the bottom part) are obtained for plane waves propagating in the $x_1$~direction (respectively in the $x_2$~direction). (a)-(c) Dispersion curves of waves propagating in $x_1$ and $x_2$ directions. The stretch ratio $\lambda_1$ is given by the darkness, as represented on the colorbar. (b)-(d) Phase velocities of $S\!H_0$ and $S_0$ at 170~Hz extracted and plotted as functions of $\lambda_1$.}
    \label{fig:expDispersionVelo}
\end{figure*}

To track systematically the anisotropy induced on both $S\!H_0$ and $S_0$ modes, a new set of measurements is performed. The point source (figure~\ref{fig:setupPlate}) is replaced by a line source to generate plane waves as shown on the left part of figure~\ref{fig:expDispersionVelo}. And the vibration is reproduced for different static stretch ratio $\lambda_1$ and frequencies.
Shaking this line source in the $(x_1,x_2)$~plane, with a 45\textdegree{} angle between the displacement and the propagation direction, allows the observation of both $S\!H_0$ and $S_0$ in one experiment. The dispersion curves obtained for different stretch ratios~$\lambda_1$ are plotted in figure~\ref{fig:expDispersionVelo}(a) (respectively \ref{fig:expDispersionVelo}(b)) for plane waves propagating in the $x_1$~direction (respectively the $x_2$~direction). The higher the stretch ratio, the darker the curves, as represented on the colorbar on the right of the figure. The behaviour depends on the propagation direction: while the slopes are increasing in the parallel direction (in the $x_1$~direction), they barely vary in the perpendicular direction (in the $x_2$~direction). Note that at the frequency $f\sim100$~Hz, an accident occurs due to a mechanical resonance of the clamp fixed to the shaker and holding the line source. This small 'anti-crossing' has no influence on the measured dispersion relation above 150~Hz.

From now on, phase velocities are extracted at an intermediate frequency of 170~Hz and plotted as a function of the stretch ratio~$\lambda_1$ in figure~\ref{fig:expDispersionVelo}(b) and \ref{fig:expDispersionVelo}(d) for the parallel and perpendicular directions (numerical values are available in Appendix~E). The velocity of $S\!H_0$ appears to vary linearly with $\lambda_1$ in the parallel direction, while this is not the case for the velocity of $S_0$. In addition, velocities of both $S\!H_0$ and $S_0$ remain almost constant in the perpendicular direction.

In this section, the experimental setup was introduced and it has allowed the observation of two in-plane modes $S\!H_0$ and $S_0$ in a soft plate with a factor of 2 between their velocities. Systematic measurements at different frequencies and stretch ratios in a highly-stretched plate allowed to follow the phase velocities changes with the stretch ratio $\lambda_1$. Those variations are also referred to as the acousto-elastic effect and is detailed in the next section.

\section*{Acousto-elastic effect\label{sec:hyperelasticity}}
The acousto-elastic effect is the change in velocities of elastic waves due to the initial stress. It is based on a non-linear elastic theory since the initial stress usually induces large deformations compared to the incremental motions describing waves propagating in a material. Here, the non-linearity first appears as a non-linear geometrical effect but we will see that a mechanical non-linearity must also be thought out when using a hyperelastic model. Numerous works have been devoted to explaining this phenomenon both theoretically and experimentally, and its main characteristics are recalled. More details can be found in Appendix~B, or in the works of Ogden, Destrade and Saccomadi~\citep{ogden_1997,destrade_2007,saccomandi_2004}. In this paper, we want to apply the acousto-elastic theory to the problem of guided waves in a plate. This will help us to understand the measured changes from previous section. Firstly, the acousto-elastic effect is used to make predictions for bulk waves. Notably, we will demonstrate the breaking of the symmetries of the elastic tensor underlying Voigt's notation in the usual study of elastic waves. We will show that this symmetry-breaking is responsible for the induced anisotropy that we have observed in figure~\ref{fig:first_obs} and~\ref{fig:expDispersionVelo}. Secondly, the acousto-elastic theory will be applied to derive the velocities of guided waves in a plate and will be compared to the experimental data.

\subsection*{Non-linear elasticity}
The constitutive law in linear elasticity, known as the Hooke's law, relates linearly the Cauchy stress~$\boldsymbol{\sigma}$ to the strain tensor~$\displaystyle \boldsymbol{\epsilon} = \nicefrac{1}{2}\left[\boldsymbol{\nabla}\mathbf{u}+(\boldsymbol{\nabla}\mathbf{u})^\mathrm{T}\right]$ where $\mathbf{u}=\mathbf{x'}-\mathbf{X}$ with $\mathbf{X}=\left(X_1,X_2,X_3\right)$ the coordinates of the natural configuration (figure~\ref{fig:setupPlate}(a)) and $\mathbf{x'}=\left({x_1}',{x_2}',{x_3}'\right)$ the set of coordinates defining the incremental motion. Using Einstein summation notation it writes:
\begin{equation}
    \sigma_{ij} = C_{ijkl}\epsilon_{kl}
    \label{eq:hooke}
\end{equation}
The link between the two second-order tensors is the fourth-order elasticity tensor $C_{ijkl}$. Combining this constitutive law with the equation of motion, {\it ie.} $\boldsymbol{\nabla} \cdot \boldsymbol{\sigma}^\mathrm{T} = \rho \,\Ddot{\mathbf{u}}$, the wave equation is obtained:
\begin{equation}
    C_{jikl} \frac{\partial^2 u_l}{\partial X_j\partial X_k} = \rho \frac{\partial^2 u_i}{\partial t^2}
    \label{eq:motion}
\end{equation}
Nevertheless, in the framework of non-linear elasticity, depending on the chosen Lagrangian or Eulerian approaches, new tensors have to be considered~\citep{ogden_1997,goriely_2017}. Here, the Cauchy stress tensor~$\boldsymbol{\sigma}$ is kept and the Green-Lagrangian strain tensor~$\mathbf{E}$ is considered. Like any strain measures, it is based on the use of the deformation gradient $\mathbf{F_s} = \mathbf{1}+\nabla\mathbf{u_s}$, where $\mathbf{1}$ is the second-order identity tensor and $\mathbf{u_s}=\mathbf{x}-\mathbf{X}$ is the static displacement (after applying an elongation in figure~\ref{fig:setupPlate}(b)) using coordinates $\mathbf{x}=\left(x_1,x_2,x_3\right)=\left(\lambda_1 X_1,\lambda_2 X_2,\lambda_3 X_3\right)$. The corresponding static strain tensor $\mathbf{E_s}$ is defined as:
\begin{eqnarray}
    \mathbf{E_s} &=& \frac{\mathbf{F_s}^\mathrm{T}\mathbf{F_s}-\mathbf{1}}{2} \nonumber\\
    &=& \frac{1}{2}\left[\nabla\mathbf{u_s} + (\nabla\mathbf{u_s})^\mathrm{T} +  (\nabla\mathbf{u_s})^\mathrm{T} \cdot\nabla\mathbf{u_s}\right]
    \label{eq:GLstrain}
\end{eqnarray}

A geometrical non-linearity is therefore evidenced as the last term in the sum. But the mechanical non-linearity must also be thought out. Typically, as introduced, it is common to use a hyperelastic law for soft media where $\boldsymbol{\sigma}$ is also a non-linear function of $\mathbf{E_s}$. This constitutive law is detailed in Appendix~B and basically relies on a strain energy density function $W$.

\subsection*{Bulk waves}
To describe waves in a pre-stressed body, an incremental approach is built as described by Ogden and Destrade~\citep{ogden_1997,destrade_2007,destrade_2012}. The main result of this theory is that all the non-linearities can be included in a new tensor $C_0$ and a wave equation similar to equation~(\ref{eq:motion}) is still obtained:
\begin{equation}
    C_{0jikl} \frac{\partial^2 u'_l}{\partial x_j\partial x_k} = \rho \frac{\partial^2 u'_i}{\partial t^2}
    \label{eq:StretchedMotion}
\end{equation} with $\mathbf{u'}(\mathbf{x},t)=\mathbf{x'}-\mathbf{x}$ an incremental displacement and $C_0$ the modified elasticity tensor that depends on $W$ and the stretch ratios~$\lambda_i$. The calculation of the tensor coefficients are also detailed in Appendix~B.
\begin{figure*}
    \centering
    \includegraphics[width=\linewidth]{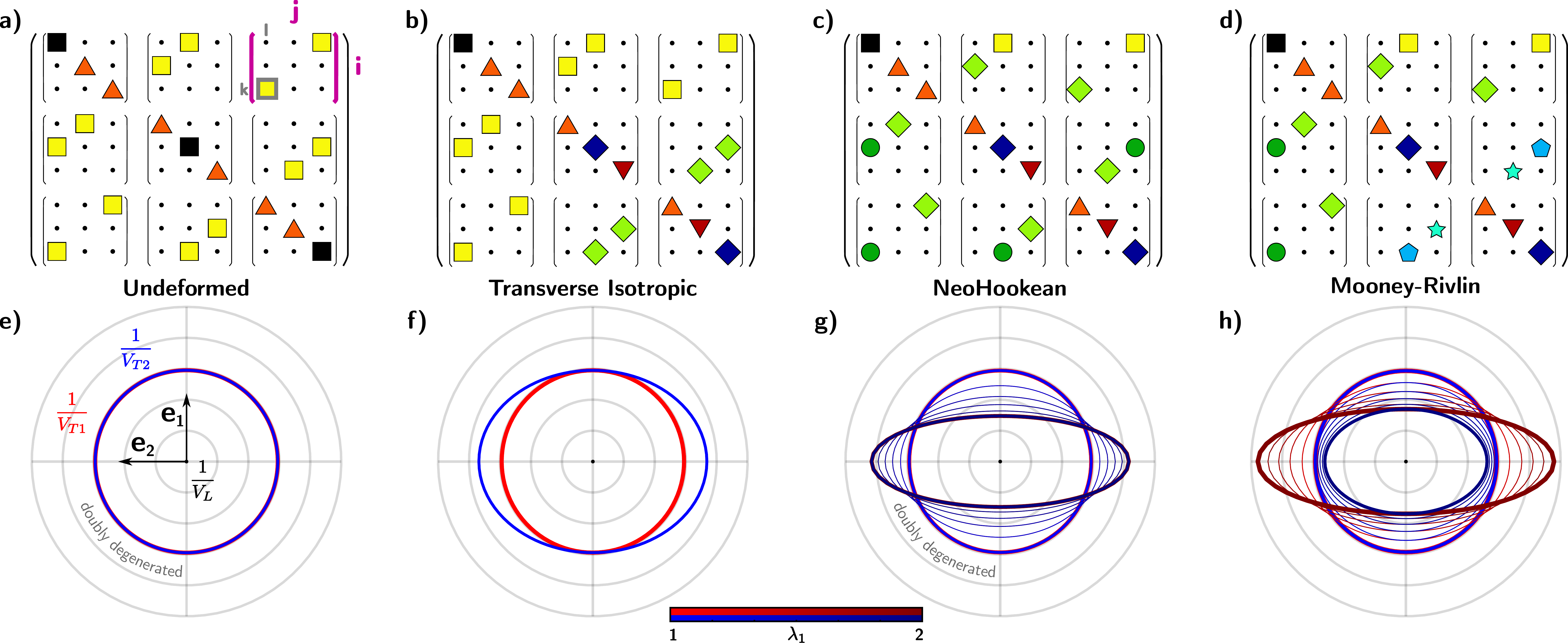}    \caption{\textbf{Elasticity tensors and slowness curves for various models assuming a uniaxial tension --} The elasticity tensor $C_{0ijkl}$ is represented as a 3x3 matrix $(i,j)$ of 3x3 matrices $(k,l)$, and the slowness curves of bulk waves propagating in the $(x_1,x_2)$~plane are plotted: $V_{T1}$ (respectively $V_{T2}$) is the velocity of the shear wave polarized in the $(x_1,x_2)$~plane (resp. along the $x_3$~axis). An equal radial spacing of 0.02 s.m$^{-1}$ is applied. First, a linear isotropic model (a,e) with $\lambda_{\text{Lamé}}=1$~GPa and $\mu=46$~kPa and a transversely isotropic solid (b,f) are considered. To plot the slowness curves in~(f), following elastic constants are used: $C_{44}/C_{66}=0.8^2$, $C_{11}=C_{22}$ and $C_{23}=C_{12}$ (Voigt notation). Then, a NeoHookean hyperelastic model (c,g) using the same mechanical constants $\lambda_{\text{Lamé}}$ and $\mu$ predicts degenerated shear waves, while a Mooney-Rivlin hyperelastic model (d,h) uses an additional constant $\alpha=0.5$. For each hyperelastic model, different stretch ratios $1\leq \lambda_1 \leq 2$ are considered.}
    \label{fig:tensors}
\end{figure*}

A major consequence of this approach is the loss of some fundamental symmetries in the $C_0$ tensor, as illustrated in figure~\ref{fig:tensors}. Indeed, in the linear approach, the elasticity tensor $C_{ijkl}$ is generally represented thanks to the Voigt notation as a 6x6 matrix, but one has to keep in mind that it contains 81 coefficients. A full representation with a 3x3 matrix $(i,j)$ of 3x3 matrices $(k,l)$ is preferred here. In this representation, in figure~\ref{fig:tensors}(a), the isotropic elastic tensor $C_{ijkl}$ has only 3 different coefficients: the two Lamé coefficients $\lambda_{\text{Lamé}}$ (orange) and $\mu$ (yellow), and a third coefficient which depends on these two constants, $\lambda_{\text{Lamé}}+2\mu$ (black). Given the fact that a stress is applied along a particular direction in our experiments, one would be tempted to consider a transverse isotropic material where $x_1$ would be the isotropy axis. The elasticity tensor for such an anisotropic medium remains as sparse as the one of the isotropic material but now contains 6 distinct coefficients as depicted in figure~\ref{fig:tensors}(b).

Let us now consider a uniaxial tension for an incompressible material $\lambda_2=\lambda_3= \lambda_1^{-0.5}$. Using the NeoHookean hyperelastic model~\citep{marckmann_2006}, the Lamé constants are also sufficient to describe the modified elasticity tensor $C_{0jikl}$ (see Appendix~B). It now contains 7 different coefficients as shown in figure~\ref{fig:tensors}(c). And, very interestingly, some fundamental symmetries are broken: $C_{0ijkl} \neq C_{0jikl}$ and $C_{0ijkl} \neq C_{0ijlk}$. However, the symmetry $C_{0ijkl} = C_{0klij}$ is preserved. Therefore, the Voigt notation is no longer valid and no usual anisotropic model can be used. The Mooney-Rivlin model brings an additional constant $\alpha$ expressing the mechanical non-linearity (see Appendix~B), and there are 9 different coefficients in the modified elasticity tensor $C_{0jikl}$ (figure~\ref{fig:tensors}(d)).

From the knowledge of the modified elastic tensor $C_{0ijkl}$, and the propagation equation~[\ref{eq:StretchedMotion}], the bulk wave velocities for any plane waves can be retrieved. Depending on the considered hyperelastic model, the two transverse waves may have degenerated velocities. Those results are summed up in the bottom part of figure~\ref{fig:tensors}. The linear isotropic, the transverse isotropic approaches and the two hyperelastic models are considered. For each model, the velocities of the three bulk waves propagating in the $(x_1,x_2)$~plane are derived and their inverse $\frac{1}{V}$, referred to as the slowness, are plotted for different values of $\lambda_1$ below the corresponding elasticity tensor in figure~\ref{fig:tensors}. In the undeformed case, circles, indicating isotropic media, are recovered. The longitudinal velocity being very large compared to the transverse one for a nearly incompressible medium, the longitudinal slowness curve appears as a single point on the graph. In the transverse isotropic model, the shear wave polarized in the $(x_1,x_2)$~plane (red in figure~\ref{fig:tensors}) remains isotropic, while the shear wave polarized along the $x_3$~axis (blue in figure~\ref{fig:tensors}) is anisotropic.
Taking into account the hyperelasticity, in the NeoHookean model, slowness curves become ellipses and shear waves are degenerated. The shear wave polarized in the $(x_1,x_2)$~plane is no longer isotropic. And with the Mooney-Rivlin model, transverse velocities are now distinct. In particular, the shear wave polarized in the $(x_1,x_2)$~plane propagates slower in the perpendicular direction in a deformed plate, as for the NeoHookean model, while the shear wave polarized along the $x_3$~axis propagates faster.

\subsection*{Hyperelastic predictions for elastic guided waves in a plate}
The case of the plate is slightly more complicated than the bulk since the solutions are no longer plane waves, but guided waves which must satisfy boundary conditions. However, the changes compared to an undeformed material are all contained in the modified elastic tensor $C_{0ijkl}$ as for the bulk case.

For the $S\!H_0$ mode, everything remains similar to the bulk and its velocity is indeed given by one of the bulk shear waves (red in figure~\ref{fig:tensors}). More details can be found in Appendices~C and D. Assuming a NeoHookean material under a uniaxial tension, it is easy to show that:
\begin{equation}
    V_{T, \parallel} = \sqrt{\frac{\mu}{\rho}}\, \lambda_1 \quad\text{and}\quad V_{T, \perp} = \sqrt{\frac{\mu}{\rho}}\, \lambda_2 
\end{equation}

The dispersion of Lamb waves in a stretched plate is more difficult to calculate~\citep{ogden_1993,mohabuth_2019}. Nonetheless, the velocity of the $S_0$ mode in the low-frequency limit has been derived by Rogerson and Fu in the equation~(3.22) of their work~\citep{rogerson_1995} for incompressible hyperelastic models. More details can be found in Appendices~C and D. Using their work and applying it to the NeoHookean model, the plate velocities write:
\begin{equation}
    V_{P, \parallel} = \sqrt{\frac{\mu}{\rho}}\sqrt{\lambda_1^2+3\lambda_3^2} \quad\text{and}\quad
    V_{P, \perp} = \sqrt{\frac{\mu}{\rho}}\sqrt{\lambda_2^2+3\lambda_3^2}
\end{equation}
At this point, it is worth recalling that the deformation does not perfectly match the one expected for a uniaxial tension. In fact, the top and bottom clamps induce a slightly different configuration where $\lambda_2=\lambda_1^{-0.41}\neq\lambda_1^{-0.5}$ as detailed earlier and in Appendix~A.

In this section, the acousto-elastic theory was detailed and it has allowed the understanding of the measured and initial stress induced anisotropy. Moreover, the equivalent elasticity tensor for incremental motions was described, bulk wave velocities were derived and the problem of guided waves in a plate was solved to obtain analytical hyperelastic predictions for the velocities of $S\!H_0$ and $S_0$. Let's compare those predictions with the aforementioned quantitative experimental measurements. 

\section*{Lack of success of the acousto-elastic theory}
The hyperelastic analytical predictions for the velocities of $S\!H_0$ and $S_0$ are superimposed on the experimental curves presented in figure~\ref{fig:expDispersionVelo}. Here we show that analytical predictions, whatever the hyperelastic model, cannot explain the observed changes, even in low stretch ratio limit. Besides, a static stress estimation is still possible and leads us to show the need for an additional physical element, that is to say the rheology of the material.

\subsection*{Comparison of experimental measurements of phase velocities of $S\!H_0$ and $S_0$ with analytical hyperelastic predictions}
\begin{figure*}
    \centering
    \includegraphics[width=\linewidth]{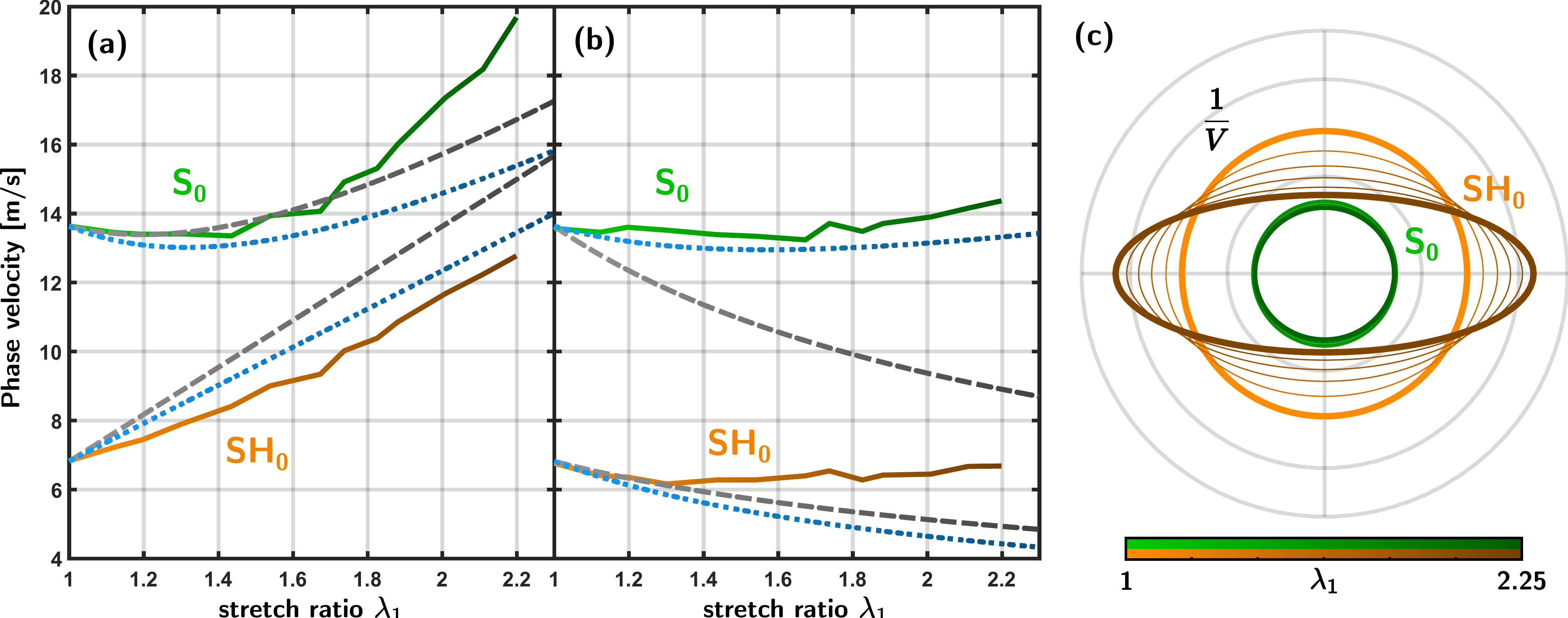}
    \caption{\textbf{Hyperelastic model predictions for velocities of $S\!H_0$ and $S_0$ --} In dashed grey lines are presented the NeoHookean model predictions, and in dotted blue lines the Mooney-Rivlin ones, in the parallel (a) and perpendicular (b) directions. Both models use the same shear modulus $\mu=46$~kPa but the Mooney-Rivlin also uses an additional constant $\alpha=0.32$. (c) Slowness curves are plotted for the Mooney-Rivlin model using previous fitting parameters.}
    \label{fig:hyperelasticVelo}
\end{figure*}

Given equations in Appendices C and D, it is straightforward to derive the equivalent elasticity tensor and to build predictions for any hyperelastic model, in order to compare them to the experimental measurements. This comparison is presented in figure~\ref{fig:hyperelasticVelo}(a) and \ref{fig:hyperelasticVelo}(b). The grey lines corresponds to the predictions using the NeoHookean model and the blue lines correspond to the Mooney-Rivlin model. The transverse velocity in the undeformed plate, at $\lambda_1=1$ directly provides $\mu=\rho {V_T}^2=46$~kPa. No fitting is needed for the NeoHookean model. However, for the Mooney-Rivlin model, $\alpha=0.32$ coefficient is obtained using a least-square procedure involving the four velocities for $1\le\lambda_1\le1.8$. For figure~\ref{fig:hyperelasticVelo}(c), the tensor of a compressible Mooney-Rivlin model was first derived using Mathematica~\citep{wolfram} and the same previous constants. Then, the slowness curves were computed using an adaptation from an open-source code\footnote{The reader should pay attention to the index convention: the one used in this open-source code is different from the one described in this paper because $C_{0ijkl} \neq C_{0jikl}$ here.}~\citep{kiefer_2022a,kiefer_2022b}. Those slowness curves should be compared to the experimental isofrequency contours displayed in figure~\ref{fig:first_obs}(d) and \ref{fig:first_obs}(h).

Although the parallel velocity of $S\!H_0$ is rightfully modelled as a linear function of the stretch ratio $\lambda_1$, it appears that none of those models can predict the slope. Furthermore, it is expected from both models that the perpendicular velocity of $S\!H_0$ decreases with $\lambda_1$ while the experimental measurements show it remains almost unchanged. Regarding these observations, we have tested various existing hyperelastic models~\citep{destrade_2010,marckmann_2006}, adding unknown mechanical constants to be determined during the least-square procedure, but none of them is able to capture both the slope of the parallel velocity and the almost unchanged perpendicular velocity of $S\!H_0$.

\subsection*{The smaller static shear modulus compared to the dynamic one tells us the importance of the frequency}
\begin{figure}
    \centering
    \includegraphics[width=.5\textwidth]{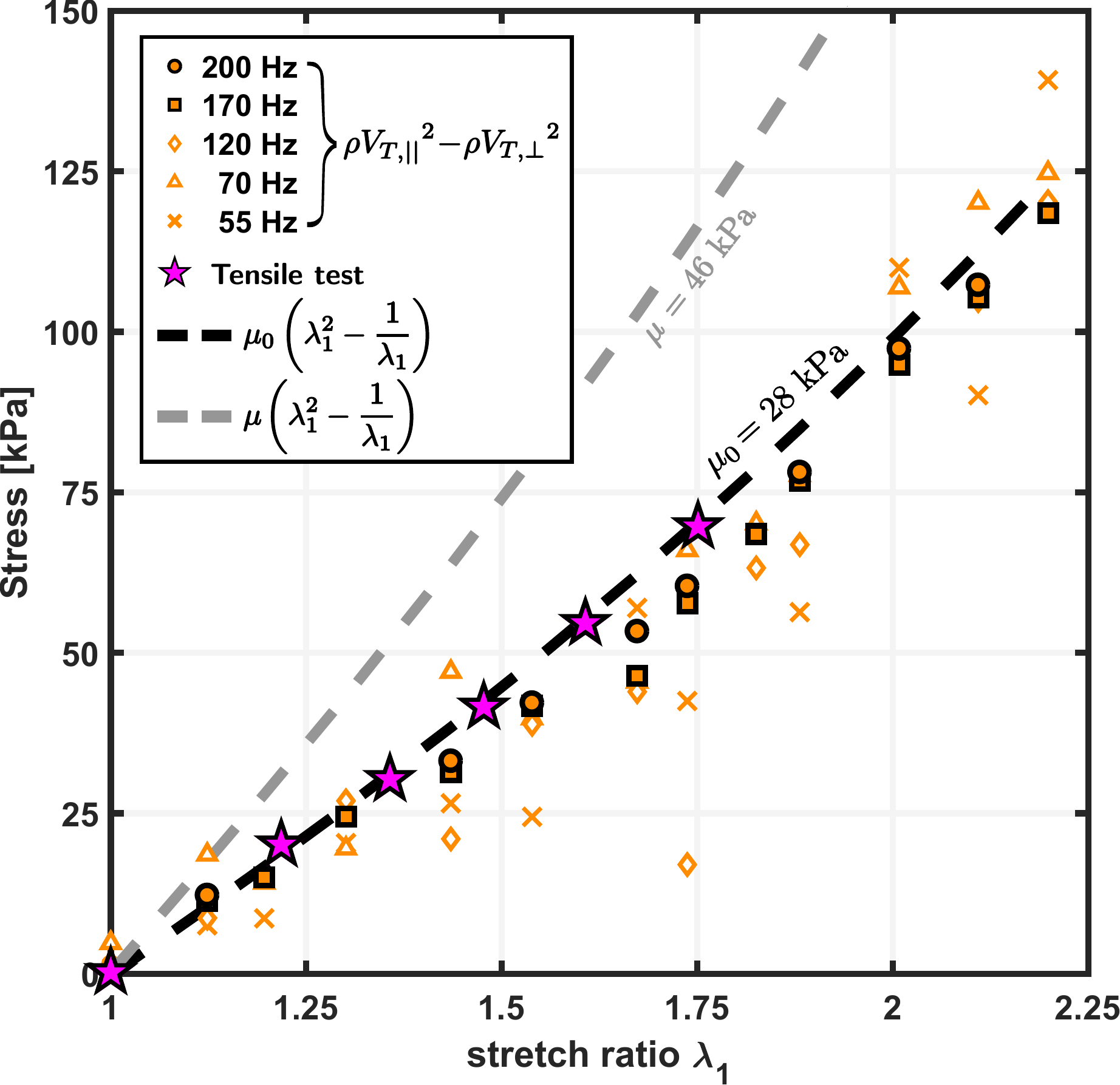}
    \caption{\textbf{Applied static stress is measured with different methods --} The observable $\rho V_{T, ||}^2 - \rho V_{T, \perp}^2$ is plotted in orange symbols for different frequencies. It is compared to an independent static measurement (magenta stars) similar to a tensile test. Two theoretical predictions assuming a NeoHookean model and a uniaxial tension are plotted in black ($\mu_0=28$~kPa) and grey ($\mu=46$~kPa) dashed line.}
    \label{fig:static}
\end{figure}

Another limit of those hyperelastic predictions appears when comparing those results to static measurements. In figure \ref{fig:static}, the static stress in the plate is plotted as a function of the measured stretch ratio $\lambda_1$ (magenta stars) during a tensile test. A static shear modulus $\mu_0=28$~kPa can be deduced using the initial slope given by the Young modulus $E=3\mu_0$. Assuming a NeoHookean model and a uniaxial tension, the stress exactly writes: $\sigma_1=\mu_0\left(\lambda_1^2-\nicefrac{1}{\lambda_1}\right)$ (since $\sigma_2=0$ is assumed) and is plotted in a black dashed line in figure~\ref{fig:static}.

This value is markedly different from the one used in figure~\ref{fig:hyperelasticVelo} ($\mu=46$~kPa) which was fixed to match the transverse velocity in the undeformed plate. This points out the influence of the frequency on the shear modulus: $\mu(\omega)\ne \mu_0$, and it needs to be taken into account. Surprisingly, this static stress can be recovered from dynamic experiments thanks to a model-independent observable, as used in~\citep{li_2022} to map stresses in thin films:
\begin{equation}
    \rho {V_{T, ||}}^2 - \rho {V_{T, \perp}}^2 = 
    \sigma_1-\sigma_2
    \label{eq:SH0squared}
\end{equation}
In figure~\ref{fig:static}, this quantity is plotted as a function of $\lambda_1$ in various orange symbols for different frequencies. It matches well with all data points and implies a very interesting feature in terms of applications: one can measure a static and local stress in a plate through dynamic perturbations~\citep{li_2022}. All the more remarkable, it seems to remain true for all the measured frequencies.

The measured static shear modulus $\mu_0$ differs from $\rho {V_T}^2$ at all frequencies. This observation highlights the frequency dependence of the shear modulus~$\mu$. In fact, this conclusion is consistent with previous work from Yasar~\citep{yasar_2013} and also with our previous works~\citep{lanoy_PNAS_2020,delory_JASA_2022} where the rheology of the used polymer has been well described by a fractional Kelvin-Voigt model:
\begin{equation}
    \mu\left(\omega\right) = \mu_0 \left[1+\left(\textrm{i}\omega\tau\right)^n\right]
    \label{eq:kelvinvoigt}
\end{equation}
with $\tau=210$~µs and $n=0.27$.
The Kelvin-Voigt model is a commonly used viscoelastic model, and its fractional derivative counterpart $\left(\textrm{i}\omega\right)^n$ with $0<n<1$ originates from so-called memory effects, where the relaxation function is given by a power-law decay as detailed in~\citep{mainardi_2010,machado_2011,meral_2010}.

In this section, we have explained the failure of the usual acousto-elastic effect to rightfully predict the phase velocities changes of $S\!H_0$ and $S_0$ modes in a soft plate. The frequency dependence of the shear modulus has been highlighted and detailed using a fractional viscoelastic model deduced from our previous rheological measurements.

\section*{Combining a hyperelastic model and the material rheology\label{sec:visco}}
From the last observations, it appears that the material rheological properties are essential in the problem and must be taken into account. To do so, the constitutive law need to be redesign to factor in both the rheology of the material when it is undeformed and its hyperelasticity when it is deformed. In this section, an experiment-inspired constitutive law is developed, visco-hyperelastic predictions for both the phase velocities $V$ and the attenuation distances $L$ are built and compared to the experimental data. All data could be fitted using our simple model.

Following the work of Destrade, Saccomandi and Ogden~\citep{destrade_2009}, the Cauchy stress tensor is rewritten as the sum of a static and a dynamic part (rather than an elastic and a dissipative part), where the static part is still given by the hyperelastic theory, but the dynamic part writes:
\begin{equation}
    \boldsymbol{\sigma}_{\textbf{dynamic}} = 2\nu\mathbf{D}+\beta\left(\mathbf{BD+DB}\right)
    \label{eq:sigmadynamic}
\end{equation}
where $\mathbf{B}=\mathbf{F}\mathbf{F}^{\rm{T}}$ is the left Cauchy-Green tensor, $\mathbf{D} = \nicefrac{1}{2}\left[\mathbf{L}+\mathbf{L}^{\rm{T}}\right]$ and $\mathbf{L}=\partial_{t} \mathbf{F} \cdot \mathbf{F}^{-1}$. Here, the importance of the fractional derivative in the $\mathbf{L}$ tensor should be pointed out. It originates from the previously depicted viscoelastic model in equation~(\ref{eq:kelvinvoigt}) and is experiment-driven. Like in a recent theoretical study~\citep{zhang_2022}, it becomes:
\begin{equation}
    \mathbf{L}=\partial_{t^n} \mathbf{F} \cdot \mathbf{F}^{-1}
\end{equation}
From this expression, it is now possible to rewrite an elasticity tensor at a given frequency, using an additional fractional viscous term:
\begin{equation}
    C_{\omega ijkl} = C_{0ijkl} + \left(\nu + \beta\,\frac{\lambda_i^2+\lambda_j^2}{2}\right)\left(\textrm{i}\omega\right)^n
\end{equation}
Considering the limit $\left(\nu,\beta\right)\rightarrow 0$, we recover the hyperelastic predictions $C_{0ijkl}$. In an undeformed plate, $(\lambda_i)\rightarrow 1$ and the rheology described in equation~(\ref{eq:kelvinvoigt}) must be recovered, leading to $\nu+\beta\!=\!\mu_0\tau^n$. This experiment-driven condition is satisfied rewriting $\beta\!=\!\beta'\mu_0\tau^n$ and $\nu\!=\!(1-\beta')\mu_0\tau^n$ so there is only one remaining unknown constant in this fractional viscous part~$\beta'$. From this, it is straightforward to derive the previous velocities using $C_{\omega ijkl}$ instead of $C_{0ijkl}$, as well as the attenuation distances $L$.

\begin{figure*}
    \centering
    \includegraphics[width=\linewidth]{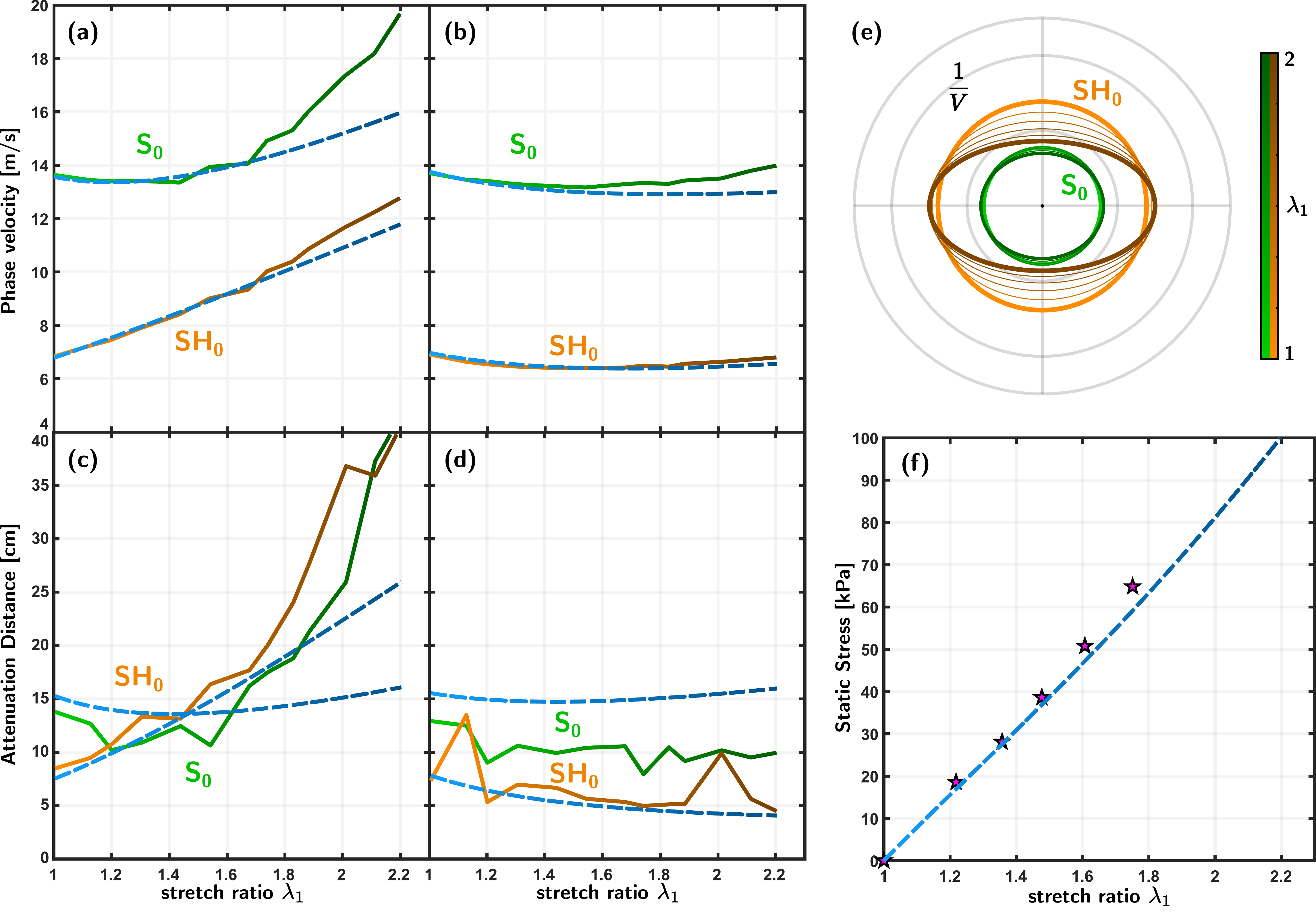}
    \caption{\textbf{Viscoelastic Mooney-Rivlin model predictions for all experimental data at 170~Hz --} The experimental measurements of the phase velocities $V$ of $S\!H_0$ and $S_0$ are plotted for the parallel (a) and perpendicular (b) directions. The measurements of the attenuation distance $L$ are added in (c) and (d). The fitting of complex wavenumbers in both directions for $1\le\lambda_1\le1.8$ provides $\alpha=0.29$ and $\beta'=0.29$. Slowness curves (e) and static stress (f) are plotted using previous fitting parameters.}
    \label{fig:viscoelasticVelo}
\end{figure*}

Doing so leads to new predictions as illustrated in figure~\ref{fig:viscoelasticVelo}. Compared to the hyperelastic predictions presented in figure~\ref{fig:hyperelasticVelo}, both the parallel (figure~\ref{fig:viscoelasticVelo}(a)) and perpendicular (figure~\ref{fig:viscoelasticVelo}(b)) velocities of $S\!H_0$ are now well captured. Same remark for $S_0$ velocities.
Now that a dissipative part has been added in the stress tensor, predictions can be made for the attenuation. The displacement maps have been studied more thoroughly and a linear regression of $\log{|\mathbf{u}(\omega)|}$ allowed a measurement of the attenuation distance $L$. It is often recognized that the attenuation is a difficult quantity to assess experimentally, especially here because the involved attenuation distances are comparable to the total propagation distance.
The fitting of the complex wavenumbers $k=\frac{\omega}{V}-iL^{-1}$ for $1\le\lambda_1\le1.8$ provides $\alpha=0.29$ and $\beta'=0.29$. Here, the phase velocities have naturally a greater weight than the attenuation distance in the least-square loss function since $\text{Re}\left[k\right] \gg \text{Im}\left[k\right]$.
Moreover, slowness curves plotted in figure~\ref{fig:viscoelasticVelo}(e) are from now on similar to the experimental spatial Fourier transform presented in figure~\ref{fig:first_obs}(d) and \ref{fig:first_obs}(h). Those corrections were possible thanks to the viscoelastic part in the Cauchy stress tensor, and particularly the frequency-dependence of the new elasticity tensor $C_{\omega ijkl}$, itself resulting from the fractional derivative model. We have also checked that the static stress predictions is still in agreement with experimental data presented in figure~\ref{fig:static}.

Regardless the distance between the experimental data for $L$ and its predictions, it is important to note that the attenuation distance of $S\!H_0$ in the parallel direction (orange curve in figure~\ref{fig:viscoelasticVelo}(c)) is multiplied by almost 4 when the stretch ratio is multiplied by 2. This originates from the combination of a geometrical and a velocity increases.
One should also remark that the phase velocities curves (figure~\ref{fig:viscoelasticVelo}(a) and (b)) are no longer well captured for a stretch ratio $\lambda\ge1.8$. In fact, a Mooney-Rivlin model remains a weakly non-linear hyperelastic model. When performing static measurements such as a tensile test, it is common to use other hyperelastic models such as Gent, Fung-Demiray or Arruda-Boyce~\citep{marckmann_2006}. Here, it is our choice to avoid adding new material constants in the Cauchy stress tensor expression especially since the dynamic part in equation~(\ref{eq:sigmadynamic}) can also become more complex and rewrites as in~\citep{destrade_2009}:
\begin{equation*}
    \boldsymbol{\sigma}_{\textbf{dynamic}} = 2\nu\mathbf{D} + \beta\left(\mathbf{BD+DB}\right) + \gamma\left(\mathbf{B}^2\mathbf{D+DB}^2\right) + \dots
\end{equation*}
Additionally, by considering the complex-evaluated elasticity tensor $C_{\omega ijkl}$, equation~(\ref{eq:SH0squared}) is no longer true and the quantity $\rho {V_{T, ||}}^2 - \rho {V_{T, \perp}}^2$ does not exactly match the applied static stress. However, knowing that $\text{Re}\left[k\right] \gg \text{Im}\left[k\right]$ and in view of figure~\ref{fig:static}, this previous quantity still remains a good estimation of the applied stress. For an exact solution, we rather suggest to use the quantity: $\displaystyle \rho \frac{\omega^2}{k_{T, ||}^2} - \rho \frac{\omega^2}{k_{T, \perp}^2}$
where $\displaystyle k=\frac{\omega}{V}-iL^{-1}$ is the corresponding complex wavenumber.

\section*{Conclusion}
In this paper, a simple experimental setup is introduced to measure the phase velocities and the attenuation of two in-plane guided modes in a plate made of a soft elastomer: the $S\!H_0$ and $S_0$ modes. In the first part, a high deformation is applied to the soft plate and the induced anisotropy is characterized. Given the initial deformation, the non-linear elastic equations can in fact be reduced to a linear equation for elastic waves provided that the elasticity tensor is modified. Notably, the underlying symmetries of the Voigt notation are broken and none of the usual anisotropic elasticity tensors are able to explain the observed anisotropy.
Then, the phase velocities changes with the stretch ratio are systematically studied and they evidence the lack of success of the usual acousto-elastic theory for soft elastomers. We demonstrate that no hyperelastic model can rightfully fit those experimental data. In fact, it is crucial to account for the rheology. Here, we introduce a fractional derivative model because we identified a fractional viscoelastic model during previous rheological measurements. This experiment-driven approach allows us to properly fit all the measured phase velocities up to an elongation of 80\% and to understand the evolution of the attenuation distances with the stretch ratio.
In the same time, static stress measurements confirm that it can approximately be estimated using the material-independent quantity $\rho {V_{T, ||}}^2 - \rho {V_{T, \perp}}^2$.
Here, a generalization of our method is accessible to other rheological models by adjusting the dynamic stress tensor. In the same way, they are of practical interest for rheological characterization of soft materials since this theoretical method captures the influence of the applied stress on the measured viscoelastic properties. Finally, our framework bridges the gap between elastic wave physics and rheology, but also paves the way for robust quantitative elastography.

\section*{Material and methods}
\subsection*{Sample preparation}
A thin plate of Ecoflex\textsuperscript{\textregistered}-0030 (Smooth-On) with dimensions $60$ cm x $60$ cm x $3$ mm is prepared by equally mixing parts A and B and a first layer is poured in the sample mould. After ten minutes, black carbon grains are deposited for displacement tracking. After 2 hours, the second layer is poured and the sample is cured for 6 hours. The rheological properties were measured with a conventional rheometer (MCR501, Anton-Paar), which operates in the plate-plate configuration.
\subsection*{Shaking}
The excitation is performed by a shaker (TIRAvib 51120, TIRA), driven by an external arbitrary wave generator (AWG 33 220, Keysight), which is itself connected to a power amplifier (analog amplifier BAA 500, TIRA).
\subsection*{Image acquisition}
The motion is captured by a charge coupled device (CCD) camera (acA4112-20um, Basler) with a $4112\times 3008$-pixel sensor, mounted with a 85-mm zoom lens (Nikon). Stroboscopic imaging is used to overcome the camera frame rate limitations and a $60$~frames movie is recorded at a given frequency.
\subsection*{Post-processing}
The in-plane displacement field is extracted using a Digital Image Correlation (DIC) algorithm and the complex monochromatic displacement is computed at the excitation frequency. With a line source, the field is coherently summed transversely to the propagation direction and a 1D spatial Fourier Transform is computed for each of the component of the displacement field and the maxima provide the phase velocities of the two modes $S\!H_0$ and $S_0$, while the slope of the linear regression of $\log{|\mathbf{u}(\omega)|}$ provides the attenuation distance.

\section*{Acknowledgments}
We are immensely grateful to Michel Destrade for his comments and the fruitful intellectual discussions. We also would like to thank Daniel Kiefer for providing and explaining how to use his code to solve for the dispersion curves of Lamb waves. This research is supported by LABEX WIFI (Laboratory of Excellence with the French Program "Investments for the Future") under Reference Nos. ANR-10-LABX-24 and ANR-10-IDEX-0001-02 PSL* and by Agence Nationale de la Recherche under Reference No. ANR-16-CE31-0015. A.D. acknowledges funding from French Direction G{\'e}n{\'e}rale de l'Armement.

\newpage
\appendix

\section*{Appendix A: Static deformation in the plate} \label{sec:appA}
\begin{figure*}
    \centering
    \includegraphics[width=\linewidth]{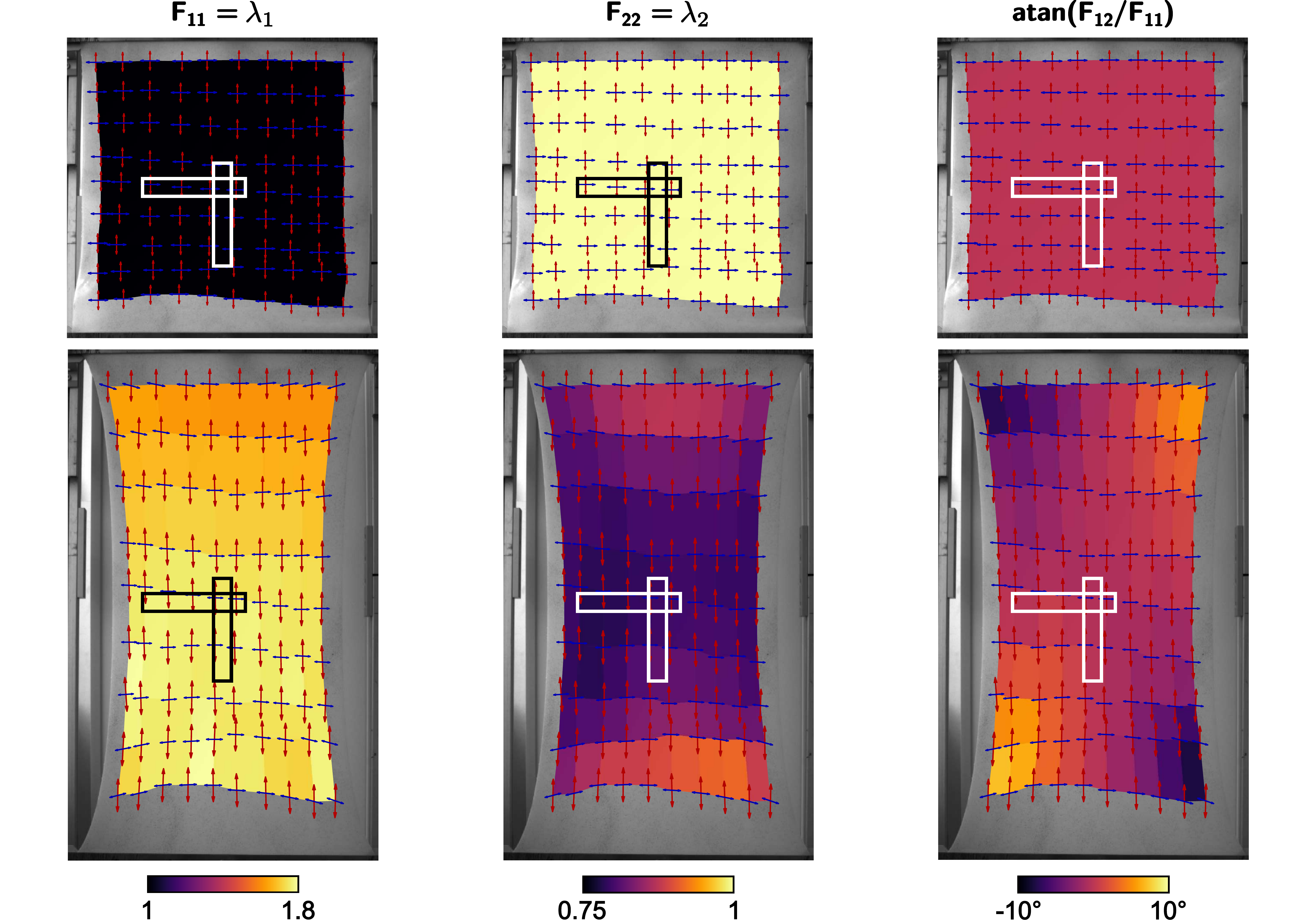}
    \caption{\textbf{Mapping the static deformation in the stretched plate --} The displacements of a mesh of black dots is extracted between the undeformed plate at rest (top) and the deformed plate submitted to a 65~kPa stress (bottom). The diagonal terms of $\mathbf{F}$ are the stretch ratios $\lambda_1$ (left) and $\lambda_2$ (center). Due to the boundaries, the angle of rotation (right) is not null everywhere. The region of interest are given by the black and white rectangles.}
    \label{fig:staticDeformation}
\end{figure*}

The same plate is stretched using a user-controlled static stress by roping weights to the bottom clamp. Here the results are shown for the undeformed plate ($\lambda_1=1$) and for a plate submitted to a stress of 65~kPa resulting in a stretch ratio of $\lambda=1.75$, in figure~\ref{fig:staticDeformation}. Some basic image processing allowed us to extract the displacement of a mesh of black dots, and the deformation gradient $\mathbf{F} = \mathbf{1}+\nabla\mathbf{u}$ is computed in the deformed plate. In figure~\ref{fig:staticDeformation}(a) (resp. (b)), the first (resp. second) element of the diagonal $F_{11}$ (resp. $F_{22}$) is plotted.
If an uniaxial tension is assumed, then $\mathbf{F}$ should be homogeneous and diagonal with $\left(\lambda_1,\lambda_2,\lambda_3\right)=\left(\lambda_1,\lambda_1^{-0.5},\lambda_1^{-0.5}\right)$ on its diagonal. However due to the boundaries, this assumption does not hold and $\mathbf{F}$ is no longer diagonal. To quantify this gap to the uniaxial tension, we have also plotted the angle $\text{atan}\left(F_{12}/F_{11}\right)$ describing the proper rotation tensor in the polar decomposition of $\mathbf{F}$ in the right part of figure~\ref{fig:staticDeformation}. When this angle is null, the deformation is purely stretching. Here we observe that the boundaries, especially in the plate corners, induce some deformations that are not purely stretching, because the top and bottom clamps are fixing the plate width to the constant initial value.
However, because of the symmetries of this static tension, $\mathbf{F}$ turns out to be almost diagonal in some regions of the plate. Namely, the centered axis $X_1=0$ and $X_2=0$ where the angle described above tends to zero. Those are the regions (white rectangles in figure~\ref{fig:staticDeformation}) where we have chosen to study in-plane guided waves and where we have measured the stretch ratios $\lambda_1$ and $\lambda_2=\lambda_1^{-0.41}$.
A last observation is this apparent vertical gradient in $F_{11}$. It can be explained since the undeformed plate is indeed not in its natural configuration, because the material is submitted to its own weight in this setup. It is well known that, due to its own weight, the deformation in the upper part of the plate will be higher than in the lower part. In fact, the bottom of the plate is exactly undeformed. Then, applying a constant static tension in the plate (by opposition to the weight which is not a constant stress) induces a higher apparent stretch ratio. This small contribution was not corrected here because it is negligible in the region of interest given by the white rectangles.

\section*{Appendix B: Hyperelastic constitutive law and incremental motions}
The hyperelastic constitutive law relies on the use of a strain energy density function~$W$ which contains all mechanical properties. It is described in the following equation:
\begin{equation}
    \boldsymbol{\sigma} = \frac{1}{J}\mathbf{F}\frac{\partial W}{\partial \mathbf{E}} \mathbf{F}^{\rm{T}} \qquad \text{with } \mathbf{E} = \frac{\mathbf{F}^\mathrm{T}\mathbf{F}-\mathbf{1}}{2} \text{ and } J=\text{det}\left(\mathbf{F}\right)
    \label{eq:hyperelasticlaw}
\end{equation}
where $\mathbf{F} = \mathbf{1}+\nabla\mathbf{u}$ is the deformation gradient, $\mathbf{u}=\mathbf{x}-\mathbf{X}$ is the displacement, $\mathbf{1}$ is the identity second-order tensor, and $\mathbf{E}$ is the Green Lagrangian strain tensor, as introduced in~\citep{ogden_1997,destrade_2007,destrade_2010,destrade_2012,saccomandi_2004}.

For an isotropic solid, $W$ is a function of principal invariants of the left ($\mathbf{B}=\mathbf{F} \mathbf{F}^{\rm{T}}$) Cauchy-Green tensor:
\begin{align*}
    I_1 &= \text{Tr}\left(\mathbf{B}\right) = \lambda_1^2 + \lambda_2^2 + \lambda_3^2\\
    I_2 &= \frac{1}{2}\left(\text{Tr}\left(\mathbf{B}\right)^2-\text{Tr}\left(\mathbf{B}^2\right)\right) = \lambda_2^2\lambda_3^2 + \lambda_1^2\lambda_3^2 + \lambda_1^2\lambda_2^2 \\
    I_3 &= \text{det}\left(\mathbf{B}\right) = \lambda_1^2\lambda_2^2\lambda_3^2 = J^2
\end{align*}
This is done to respect the invariance of $W$ under a permutation of $\left(\lambda_1,\lambda_2,\lambda_3\right)$.\\
To describe waves in a pre-stressed body, an incremental approach is built as described by Ogden and Destrade~\citep{ogden_1997,destrade_2007,destrade_2012} and the equation of motion still stands for an incremental displacement $\mathbf{u'}(\mathbf{x},t)=\mathbf{x'}-\mathbf{x}$, but with a new elasticity tensor~$C_0$:
\begin{equation}
    C_{0jikl} \frac{\partial^2 u'_l}{\partial x_j\partial x_k} = \rho \frac{\partial^2 u'_i}{\partial t^2}
    \label{eq:motion}
\end{equation}
where the coefficients of the tensor $C_0$ are expressed as:
\begin{align}
    C_{0iijj} &= \frac{\lambda_i\lambda_j}{J} \, W_{ij}& \nonumber\\
    C_{0ijij} &= \frac{\lambda_i^2}{J} \, \frac{\lambda_i W_i - \lambda_j W_j}{\lambda_i^2-\lambda_j^2} &(i\neq j, \lambda_i\neq\lambda_j) \nonumber\\
    C_{0ijij} &= \frac{C_{0iiii}-C_{0iijj}+\lambda_i W_i/J}{2} &(i\neq j, \lambda_i=\lambda_j) \label{eq:elastictensor}\\
    C_{0ijji} &= \frac{\lambda_i\lambda_j}{J} \, \frac{\lambda_j W_i - \lambda_i W_j}{\lambda_i^2-\lambda_j^2} &(i\neq j, \lambda_i\neq\lambda_j) \nonumber\\
    C_{0ijji} &= \frac{C_{0iiii}-C_{0iijj}-\lambda_i W_i/J}{2} &(i\neq j, \lambda_i=\lambda_j) \nonumber
\end{align}
where $\displaystyle W_i = \frac{\partial W}{\partial \lambda_i}$ and $\displaystyle W_{ij} = \frac{\partial^2 W}{\partial \lambda_i \partial \lambda_j}$.\\
Using an incompressible material involves some modifications. The strain energy density function~$W$ writes only as a function of invariants $I_1$ and $I_2$, since $J\rightarrow1$. Here two incompressible models are presented: the simplest and physically-based NeoHookean model described in equation~[\ref{eq:neohookean}], and the commonly used Mooney-Rivlin model described in equation~[\ref{eq:mooneyrivlin}].
\begin{eqnarray}
    W_\text{NH} &=& \frac{\mu}{2}\left(I_1-3\right)
    \label{eq:neohookean} \\
    W_\text{MR} &=& \frac{\mu}{2}\Big[(1-\alpha)(I_1-3)+\alpha(I_2-3)\Big]
    \label{eq:mooneyrivlin}
\end{eqnarray}
Note that the NeoHookean model is the special case of the Mooney-Rivlin for $\alpha=0$.
And the incompressible condition is taken into account by implementing a Lagrange multiplier $p$ so that equation~[\ref{eq:hyperelasticlaw}] is rewritten:
\begin{equation*}
    \boldsymbol{\sigma} = -p\mathbf{1} + \mathbf{F}\frac{\partial W}{\partial \mathbf{E}} \mathbf{F}^{\rm{T}}
\end{equation*}
For a Mooney-Rivlin solid, the Cauchy stress tensor simplifies:
\begin{equation*}
    \boldsymbol{\sigma} = -p\mathbf{1}+\mu\Big[(1-\alpha)\mathbf{B}-\alpha\mathbf{B}^{-1}\Big]
    \end{equation*}
The equations for the elasticity tensor~$C_0$ remain valid using an incompressible hyperelastic model but the equation of motion~[\ref{eq:motion}] is rewritten and longitudinal waves are no longer solutions:
\begin{equation*}
    C_{0jikl} \frac{\partial^2 u'_l}{\partial x_j\partial x_k} -\frac{\partial p'}{\partial x_i} = \rho \frac{\partial^2 u'_i}{\partial t^2} \label{eq:appB}
\end{equation*}
with $p'$ the incremental Lagrange multiplier.

\section*{Appendix C: $S\!H_0$ and $S_0$ velocities predictions using a hyperelastic constitutive law in an incompressible approach}
It is possible to solve analytically equation~\ref{eq:appB} for bulk waves~\citep{destrade_2007}, and we have seen that the $S\!H_0$ velocity directly equals to the corresponding bulk wave, and writes for a propagation direction $i$:
\begin{align*}
    \rho {V_{T, i}}^2 &= C_{0ijij} = \lambda_i^2 \,\frac{\lambda_i W_i-\lambda_j W_j}{\lambda_i^2-\lambda_j^2}
\end{align*}
When the plate is subjected to an uniaxial stress along the $x_1$ axis, one should consider $(i=1, j=2)$ for the parallel velocity $V_{T, \parallel}$, and $(i=2, j=1)$ for the perpendicular one $V_{T, \perp}$.\\
It is also possible to solve equation~\ref{eq:appB} for a plate geometry, and to obtain analytically the velocity of the $S_0$ mode in the low-frequency limit~\citep{rogerson_1995}, also referred to as the plate velocity, and it writes for a propagation direction $i$:
\begin{align*}
    \rho {V_{P, i}}^2 &= C_{0i3i3}+3C_{03i3i} = \left(\lambda_i^2+3\lambda_3^2 \right)\,\frac{\lambda_i W_i-\lambda_3 W_3}{\lambda_i^2-\lambda_3^2}
\end{align*}
When the plate is subjected to an uniaxial stress along the $x_1$ axis, one should consider $(i=1)$ for the parallel velocity $V_{P, \parallel}$, and $(i=2)$ for the perpendicular one $V_{P, \perp}$.

\section*{Appendix D: $S\!H_0$ and $S_0$ velocities predictions using a hyperelastic constitutive law in a compressible approach}
Numerically, we have decided to work with a compressible material, keeping the value of $\lambda_{\text{Lamé}}=1$~GPa as a material constant, since it has also permitted plotting the slowness curves. To do so, it is necessary to use a modified strain energy density function with a volumetric part.
\begin{align*}
    W_\text{NH compressible} &= \frac{\mu}{2}\left(\frac{I_1}{J^{2/3}}-3\right) + \frac{K}{2}(J-1)^2 \\
    W_\text{MR compressible} &= \frac{\mu}{2}\Bigg[(1-\alpha)\left(\frac{I_1}{J^{2/3}}-3\right)+\alpha\left(\frac{I_2}{J^{4/3}}-3\right)\Bigg] + \frac{K}{2}(J-1)^2
\end{align*}
where $\displaystyle K=\lambda_{\text{Lamé}}+\frac{2}{3}\mu$ is the material bulk modulus.\\
Similarly, solving equation~\ref{eq:StretchedMotion} for bulk waves is easy and the $S\!H_0$ velocity directly writes for a propagation direction $i$:
\begin{align*}
    \rho {V_{T, i}}^2 &= C_{0ijij} = \frac{\lambda_i^2}{J} \,\frac{\lambda_i W_i-\lambda_j W_j}{\lambda_i^2-\lambda_j^2}
\end{align*}
When the plate is subjected to an uniaxial stress along the $x_1$ axis, one should consider $(i=1, j=2)$ for the parallel velocity $V_{T, \parallel}$, and $(i=2, j=1)$ for the perpendicular one $V_{T, \perp}$.\\
To derive analytically the plate velocity, it is again possible to use Rogerson's work~\citep{nolde_2004,rogerson_2009} and it writes for a propagation direction $i$:
\begin{equation*}
    \rho V^2 = C_{0iiii}-{C_{0ii33}}^2/C_{3333}
\end{equation*}
In order to take the incompressible limit to get back on equation from Appendix~C, one should combine this expression with the free boundary condition $\sigma_3=0$. This last provides a non-zero value for $\lambda_{\text{Lamé}}\left(J-1\right)$ so that $J\ne 1$ actually, even if the material is assumed incompressible. Then, this value of $J$ is re-injected in the previous expression of $\rho V^2$ to have back the proper incompressible limit for the plate velocity.\\
It is also possible to solve numerically for the Lamb dispersion curves using the full elasticity tensor by carefully replacing $J$. This numerical approach was in particular developed to derive the slowness curves of the $S_0$ mode.

\section*{Appendix E: Experimental data for the phase velocities and attenuation distances of $S\!H_0$ and $S_0$ at 170~Hz, used in fitting procedures}

\begin{table*}[h]\centering\footnotesize
\caption{Phase velocities $V$ (m/s) and attenuation distances $L$ (cm) of $S\!H_0$ and $S_0$ modes in parallel and perpendicular directions as functions of the stretch ratio $\lambda_1$ at 170~Hz.}
\begin{tabular}{l|rrrrrrrrrrrrr}
$\lambda_1$ & 1 & 1.123 & 1.196 & 1.301 & 1.435 & 1.539 & 1.673 & 1.737 & 1.825 & 1.881 & 2.008 & 2.109 & 2.199 \\
\hline
$\displaystyle V_{T, \parallel}$ & 6.83 & 7.23 & 7.44 & 7.90 & 8.41 & 9.01 & 9.34 & 10.02 & 10.39 & 10.86 & 11.68 & 12.24 & 12.77 \\
$\displaystyle V_{T, \perp}$ & 6.72 & 6.46 & 6.36 & 6.27 & 6.22 & 6.22 & 6.23 & 6.30 & 6.27 & 6.38 & 6.45 & 6.53 & 6.61\\
$\displaystyle V_{P, \parallel}$ & 13.64 & 13.45 & 13.40 & 13.41 & 13.35 & 13.94 & 14.07 & 14.92 & 15.31 & 16.02 & 17.36 & 18.19 & 19.69\\
$\displaystyle V_{P, \perp}$ & 13.52 & 13.27 & 13.23 & 13.10 & 13.03 & 12.98 & 13.10 & 13.15 & 13.1 & 13.24 & 13.32 & 13.60 & 13.80\\
\hline
$\displaystyle L_{T, \parallel}$ & 8.69 & 9.68 & 10.95 & 13.53 & 13.36 & 16.59 & 17.89 & 20.26 & 24.24 & 27.92 & 36.03 & 36.13 & 40.78 \\
$\displaystyle L_{T, \perp}$ & 7.26 & 13.42 & 5.25 & 6.88 & 6.58 & 5.55 & 5.24 & 4.8 & 5.02 & 5.08 & 9.83 & 5.54 & 4.38\\
$\displaystyle L_{P, \parallel}$ & 13.99 & 12.88 & 10.40 & 11.11 & 12.66 & 10.85 & 16.41 & 17.71 & 19.01 & 21.52 & 26.16 & 37.51 & 41.76\\
$\displaystyle L_{P, \perp}$ & 12.85 & 12.42 & 8.94 & 10.53 & 9.85 & 10.33 & 10.48 & 7.87 & 10.38 & 9.09 & 10.10 & 9.43 & 9.87\\
\end{tabular}
\end{table*}

\newpage
\bibliographystyle{elsarticle-num-names}

\end{document}